\documentclass[aps,prb,twocolumn,superscriptaddress,oneside,floatfix,amsmath,showpacs,amssymb]{revtex4}
\usepackage{graphicx}
\usepackage{dcolumn}
\usepackage{bm}
\usepackage[usenames]{color}
\usepackage{amsmath}
\usepackage{hyperref}

\newcommand{\dd}{\text{d}}

\begin{document}

\bibliographystyle{unsrt}

\title{Slow quenches in XXZ spin-chains -- the role of Galilean invariance breaking}

\author{Piotr Chudzinski}
\email{P.M.Chudzinski@uu.nl}
\affiliation{Institute for
Theoretical Physics, Center for Extreme Matter and Emergent
Phenomena, Utrecht University, Leuvenlaan 4, 3584 CE Utrecht, The
Netherlands}

\date{\today}

%%%%%%%%%%%%%%%%%%%%%%%%%%%% ABSTRACT

\begin{abstract}
We study a XXZ spin-chain in a gapless Tomonaga-Luttinger liquid
(TLL) phase with time dependent anisotropy of spin exchange
interactions. To begin we focus on a linear ramp of $J_z$,
starting at XX point and slowly increasing towards the
anti-ferromagnetic Heisenberg point. Although the problem of a
linear ramp in the TLL has been recently under intense scrutiny in
a perturbative \emph{g-ology} framework, an aspect that has been
overlooked so far is the role of the Galilean invariance breaking.
We find that, although the differential equation that needs to be
solved to find time evolution of the system is substantially more
complicated, in some cases exact analytic solutions can be given.
We obtain them for the linear ramp in the limit of small $J_z$ as
well as $J_z\rightarrow 1$, and for such protocols that are
tailored to keep the Galilean invariance breaking term constant
for every $J_z$. We point out the features of dynamics during the
quench that stays unaltered, and those that need to be taken with
care when Galilean invariance breaking is present. We are able to
deduce that it is the shape of the propagating front that is
affected in the most pronounced way.
\end{abstract}

%\pacs{}

\maketitle

%%%%%%%%%%%%%%%%%%%%%%%%%%%% INTRODUCTION

\section{Introduction}

In recent years a non-equilibrium dynamics of a quantum system subjected to a non-adiabatic variation of its parameters, has grown into a quite important field of research\cite{Gring-12,Polkovnikov-11}. This is thanks to ground-breaking advances in simulations and experiment\cite{Trotzky-12}, which has allowed us to probe\cite{Kollath-06} intriguing, non-perturbative phenomenology of the quenched system\cite{Gritsev-07prl}. As it frequently happens, the 1D systems plays a prominent role\cite{Gritsev-06,PolkovnikovGritsev08}, because exact analytical solutions are available for also for a strongly correlated regime\cite{Cazalilla06,IucciCazalilla09}. It should be emphasized that any realistic experimental implementation of the quench has to be extended over some finite time, say from $t=0$ to $t=\tau$, which has raised an interest in a dynamics of a quantum system \emph{during} the quench\cite{Dora-11}. This is a subject of our study.

The common description\cite{CalabreseCardy06} of a dynamics of a 1D system that has undergone a quench is given in terms of propagating front. That is
entangled particles propagate throughout the system and carry
information about the change of the hamiltonian. It clearly
manifests in correlation functions of the system where, in
experimental\cite{Cheneau-12}, numerical\cite{Salvatore-fronts} and analytical\cite{moj-quench} studies, one clearly observes
different behaviour inside and outside the light cone, with a
front that is moving with an instantaneous velocity of collective eigen-particles of the
system. This intuitive picture has a hidden assumption that the
physics obeyed by the moving particles that propagate the signal
do not depend on their movement, it is an assumption of the
Galilean invariance. It is then crucial to ask the following
question: up to what extend this simple picture is valid when
Galilean invariance is broken?

The question we have raised is not of a pure academic interest. One of the most important models in 1D physics, the spin XXZ model does not obey Galilean invariance.
Its high prominence in the field of low dimensional systems is because it is solvable (using the Bethe-ansatz techniques) hence the TLL parameter $K$ and velocity $u$ are known exactly. Also numerical methods to study this model have a long history and are very well developed\cite{Pozsgay-14,Bonnes-num-prl14}. Other models have been mapped on XXZ model, to mention two leg ladder with strong on-rung coupling as an example\cite{Giam-ladd-to-XXZ}. Finally, several experimental realization of this model has been proposed, mostly based on 1D Mott insulators\cite{Essler-03,Bouillot-prl08}, but also in cold atoms systems\cite{Trotzky-08}. The problem under consideration has been already under scrutiny, in Ref.\cite{Pollmann-XXZ-perturb}, where a combination of numerical and perturbative (\emph{g-ology}) method was used. A systematic discrepancy between analytical solution and numerics was found, especially in the area of the propagating front, when the final $J_z$ ($\equiv\Delta$) was increasing. We hope to be able to understand this effect with our exact analytic solutions.

Gapless phases are probably the most interesting from the point of view of time-dependent quench dynamics because there is no gap that would inhibit propagation of the lowest energy  excitations appearing upon quenching\cite{PolkovnikovGritsev08,Collura-15}. A range of parameters where XXZ model is in the gapless TLL phase is quite broad, hence from the experimental/numerical viewpoints there is a space to vary the model's parameters. It should  be emphasized that from the exact analytic solution we know not only the precise values of $K, u$ but also non-universal amplitudes of each correlation term\cite{Caux-03prb,Collura-15}. If we assume that this functional dependence stays the same also during the quench, then this paves the way for a parameter free comparison with high precision numerical and experimental studies\cite{Trotzky-08}. This could in principle allow to investigate deviations\cite{Essler-14} from non-equilibrium TLL predictions, for instance due to irrelevant operators that may introduce thermalization, or help to resolve recently raised issue of the validity of generalized Gibbs ensemble in this model\cite{Pozsgay-14, Prosen-XXZ-15}. Hence, the existence of an exact analytical solution is of an uttermost importance. Recently, analytical solutions for several various quenching protocols have been found in Ref.\cite{moj-quench}. All were for the case when the Galilean invariance is obeyed, which obviously raises a question if analogous results may be provided when the Galilean invariance is broken.

\section{Model}

\subsection{Time dependent TLL}

We study a 1D system in a gapless Tomonaga-Luttinger liquid (TLL) phase with a time dependent parameters. The TLL hamiltonian reads\cite{Haldane-1Dmain, giamarchi_book_1d}:

\begin{equation}\label{eq:ham-TLL-def}
    H = \int \frac{dx}{2\pi}
    \left[(u(t)K(t))(\pi \Pi(x,t))^{2}+\left(\frac{u(t)}{K(t)}\right)(\partial_{x} \phi(x,t))^{2}\right]
\end{equation}

where $u(t),K(t)$ are velocity and TLL
parameter of collective bosonic mode, these
depend on underlying theory. The
Eq.\ref{eq:ham-TLL-def} is written in terms of density fields $\phi(x)$ and
canonically conjugate fields $\theta(x)$, with $\Pi(x)=\partial_{x}\theta(x)$. To define them,
first one considers a 1D theory of spinless fermions $c^{\dag}$ and extracts their long wavelength behavior around the Fermi
points given by the fields $\psi(x)$: $c^{\dag}(x)=\exp(i k_F
x)\psi_{R}^{\dag}(x)+\exp(-i k_F
x)\psi_{L}^{\dag}(x)$. Then introduces
the bosonic fields, the collective modes,
$\psi_{L,R}(x)=\kappa_{R,L}\frac{1}{2\pi\alpha}\exp(i[\sum(\phi_{L,R}(x)\pm\theta_{L,R}(x))])$
(where $\kappa_{R,L}$ is a constant operator,
Majorana fermion, introduced to ensure proper anti-commutation
relations). In our problem the bosonic fields are time dependent and obey the following differential equation which describes the time evolution of the system:
\begin{equation}
\frac{d\phi(q,t)}{d t} = u(t)K(t)q \theta(q,t)
,\frac{d\theta(q,t)}{d t} = -\frac{u(t)}{K(t)} q \phi(q,t)
\end{equation}
Following Ref.\cite{Bernier14,DziarmagaTylutki11} we introduce an auxiliary function $F(t)$ which is \emph{implicitly} defined in the following way:
\begin{equation}
\phi(q,t)= 2\sqrt{\pi K_0}|q|(a_q F^*(t)+ a_{-q}^{\dag}F(t))
\end{equation}
\begin{equation}
\theta(q,t)= \frac{1}{q u(t)K(t)}\sqrt{\frac{\pi K_0}{2|q|}}(a_q
\partial_t F^*(t)+ a_{-q}^{\dag}\partial_t F(t))
\end{equation}
where $a_q$ are bosonic operators that diagonalize the hamiltonian
at $t=0$. The time evolution of the system is encapsulated inside
$F(t)$. We write down a single ODE which is solved by $F(t)$:
\begin{equation}
\frac{\dd^2}{\dd t^2}F(q,t)+\partial_t
Log[u(t)K(t)]\frac{\dd}{\dd t} F(q,t)+u^2(t) q^2 F(q,t)=0,
\label{eq:DGLa-XXZ}
\end{equation}

At this moment we have generalized the result of Ref.\cite{Bernier14}. When
Galilean invariance is preserved, as it was assumed in Ref.\cite{Bernier14}, then
$uK=cste$ and the second term in Eq.\ref{eq:DGLa-XXZ} drops. We abandon this
assumption and arrived at a more general differential equation in
a form $\frac{\dd^2}{\dd t^2}F(q,t)+\theta_{1}q\frac{\dd}{\dd t}
F(q,t)+\theta_{2}q^2 F(q,t)=0$ which, is much more difficult to
solve when $\theta_1 \neq 0$. In the following we shall point out
a few cases for which the solution $F(t)$ of Eq.\ref{eq:DGLa-XXZ} in a closed
analytic form exists.

\subsection{XXZ model}

As already mentioned in the introduction, one prominent example
where the Galilean invariance is not preserved is the XXZ
spin-chain defined by the following hamiltonian:
\begin{equation}\label{eq:spinXXZ-def}
H_\text{XXZ}(t)=J\sum_i\bigl[S_i^xS_{i+1}^x+S_i^yS_{i+1}^y+\Delta(t)\,S_i^zS_{i+1}^z\bigr]
\end{equation}
The Jordan-Wigner transformation allows to re-express the spin
problem in terms of the spinless fermion fields $\psi(x)$. For $|\Delta|<1$ the
spectrum of the system is gapless and its low energy dynamics can
be expressed using the TLL hamiltonian, Eq.\ref{eq:ham-TLL-def}. The $\Delta=0$ is a point where fermions are non-interacting and $K=1$, while at $\Delta=1$ (and $J>0$) a gap opens in the spectrum that corresponds to an onset of an anti-ferromagnetic order. Below we are going to take initial $\Delta(t=0)=0$, although our results can be straightforwardly generalized to any initial/final $|\Delta(t)|<1$. For XXZ model the TLL parameters are non-perturbative and known exactly:
\begin{equation}\label{eq:v-in-XXZ}
  u=\frac{J\pi}{2}\frac{\sqrt{1-\Delta(t)^2}}{\arccos(\Delta(t))}
\end{equation}
\begin{equation}\label{eq:K-in-XXZ}
  K=\frac{\pi}{2}\frac{1}{\pi-\arccos(\Delta(t))}
\end{equation}

one directly checks that $uK \neq cste$. In the following we
consider a time dependent variation $\Delta(t)$. By substituting
Eq's.\ref{eq:v-in-XXZ}-\ref{eq:K-in-XXZ} to Eq.\ref{eq:DGLa-XXZ} this translates to
the following temporal variations of the new coefficient of the
differential equation:
\begin{multline}\label{eq:theta-depend}
  \theta_1=\partial_t
Log[u(t)K(t)]=\\
\left[\frac{\frac{\pi ^2 \Delta }{4}-\arcsin(\Delta ) \left(2 \sqrt{1-\Delta ^2}+\Delta\arcsin(\Delta )\right)}{\left(\Delta ^2-1\right) \left(\pi -\arccos(\Delta )\right) \arccos(\Delta )}\right]\partial_t \Delta(t)
\end{multline}
and $\theta_2$ is simply a square of Eq.\ref{eq:v-in-XXZ}.
These coefficients, given by a complicated formula involving inverse trigonometric functions, do not allow for any further analytic treatment. We need to look for a good approximate formulas. It turns out that $u(\Delta)$ can be quite well approximated by
$J\sqrt{1+\Delta}$, a formula that is nearly the same as the one
for a velocity in a TLL with the Galilean invariance preserved, where
$u=V_F\sqrt{1+2g}$ (albeit factor two of a difference). We then
conclude that the position of the front during the quench should
be the same in both cases. As for the other term given by Eq.\ref{eq:theta-depend}, the factor in
front of a derivative $\partial_t \Delta(t)$, has a singularity as
we approach the point $\Delta=1$, where a transition of XXZ model
to a gapped phase takes place. We can then immediately distinguish
two substantially different cases: \textbf{A)} a derivative $\partial_t \Delta(t)$ stays finite during the quench \textbf{B)} a
quench is continuous at $t=\tau$, hence a derivative $\partial_t \Delta(t)=0$ at  $t=\tau$ (when quenched is finished). In the case \textbf{A)}, the new term in Eq.\ref{eq:DGLa-XXZ} can
grow up to very large values and dominate. In the case \textbf{B)} this
term either stays constant or goes down to zero even if the system reaches the point $|\Delta|=1$. Finally, we note that the above described distinct behaviour of $u(\Delta)$, that suggest $g_2-g_4 \rightarrow 0$, and the divergent $\theta_1$ term, which is proportional to $g_2-g_4$ in the perturbative g-ology language, naively poses a paradox. However this unusual combination of coefficients is simply a manifestation of the fact that the perturbative g-ology language and cannot be used for spin chains at strong $\Delta$. In particular perturbative framework cannot be used for time dependent problems when this issue explicitly enters into coefficients of the central ODE, Eq.\ref{eq:DGLa-XXZ}.

%we will refer to it throughout when we wish to put our findings on firm ground.

%We take a model where one of the hamiltonian's parameters vary in time. To get the time evolution of the system, we follow a standard procedure: write the time-dependent eigenstates as a linear combination.

%In general we arrive at the following equation

%The novelty, in comparison with BCKO is that we do not assume , which is equivalent of saying that %we abandon the condition of Galilean invariance of the system during the quench. Now we wish to %explore the consequences of the Galilean invariance breaking.

\section{Linear ramp}

%\subsection{Toy model (shall I put it?????)}

%We begin with the simplest case when a coefficient next to the new
%term is time-independent. The equation for $F(t)$ now reads:

%where we take a $\Delta(t)$ that depends linearly on time,
%effectively the same $u(t)$ like in Ref.BCKO. As for $d Log uK(t)/
%d t = cste$, it is questionable if such situation may indeed
%occur, as it corresponds to exponentially increasing (in time)
%term next to momentum operator $\Pi(x)$ in the hamiltonian, that
%is no current conservation, hence a quasi-open system coupled
%in-elastically to a momentum absorbing bath. %In perturbative
%language this is exponentially increasing difference between g_2
%and g_4 while keeping velocity dependent only on linear term \sim
%g_2.
%Nevertheless, we decided to introduce such protocol because
%it allows to identify a difference caused by Galilean symmetry
%breaking in a particularly clear manner, since the solution reads:

\subsection{Small $\Delta$}

For small $\Delta$ we find that the
$\theta_1$ term depends linearly on $\Delta$, that is
$Log[u(t)K(t)]\approx
\left(\frac{4}{\pi^2}-\frac{1}{2}\right)\Delta(t) $, where we took
a Taylor series expansion of logarithm function around $Log(1)=0$. If we further assume the simplest time-resolved protocol, the linear ramp: $\Delta(t)=\Delta t/\tau$ then the differential equation reads:
\begin{equation}\label{eq:XXZ-Heisen-ODE}
(J q)^2 \left(1+\frac{\Delta  t}{\tau }\right) F(t)-\theta \frac{\Delta  t}{\tau }F'(t)+F''(t)=0
\end{equation}
where $\theta=\left(-\frac{4}{\pi^2}+\frac{1}{2}\right)$. It is possible to give an analytic solution for such equation:
\begin{widetext}
\begin{equation}\label{eq}
  F(t)= e^{\frac{J^2 q^2 t}{\theta }}\left( c_1 H_{\frac{J^2 q^2 \tau (J^2 q^2 +\theta ^2) }{\Delta  \theta ^3}} \bigg[ \kappa_1 \bigg]+c_2 1F1\left[-\frac{J^2 q^2 \tau (J^2 q^2 +\theta ^2) }{\Delta  \theta ^3},\frac{1}{2},\kappa_1^2\right]\right).
\end{equation}
where the variable $\kappa_1=\frac{\sqrt{\Delta \theta} t}{\sqrt{2\tau }}-\frac{\sqrt{2\tau } J^2 q^2 }{\sqrt{\Delta \theta ^3}}$. The $H[\kappa]$ is the Hermite polynomial and $1F1$ is the hypergeometric function. When the sign of $\Delta(t)$ does not change during the quench, this solution can be also expressed in terms of the Bessel functions:
\begin{equation}\label{eq:XXZ-linear-Bessel}
  F(t)=\frac{\Upsilon_2^{3/2}}{6(\Delta\tau^2q^2J^2)^{1/3}}{\rm e}^{\frac{\Delta
\,t\theta}{2\tau}}\left[{\it C1}\sqrt{3}{\sl J}_{1/3}\left(
\frac{\left(\Upsilon_2 \right)^{3/2}}{12{J}^{2}{q}^{2}{\tau}^{2}\Delta}\right)+
{\it C2}\sqrt{3}{\sl J}_{-1/3}\left(\frac{ \left( \Upsilon_2 \right)^{3/2}}{12{J}^{2}{q}^{2}{\tau}^{2}\Delta}\right)\right]
\end{equation}

\end{widetext}

where, this time the variable $\Upsilon_2=4\,\Delta\,{J}^{2}{q}^{2}t\tau+4\,{J}^{2}{q}^{2}{\tau}^{2}-
{\theta}^{2}{\Delta}^{2}$ was chosen.
It is indeed quite easy to compare Eq.\ref{eq:XXZ-linear-Bessel} with the solution given by
Ref.\cite{Bernier14}. We immediately see two effects caused by the new term:
an exponential damping factor in front of $F(t)$ and a shift of
an argument.  The first effect is straightforward to
interpret as it resembles a damping of a harmonic oscillator. All correlation functions, which are proportional to momentum integral of the $F(q,t)$, will acquire this extra exponential. Although for spin-chain, where $\theta$ coefficient is tiny, this factor stays close to one, it is still remarkable as it manifest an emergence of a characteristic time-scale present in a time dependent problem. Moreover the time and space are not any longer equivalent. Furthermore, for the correlation functions of $\theta(x)$ field that are proportional to $(\partial_t F(q,t))^2$, this new time dependent factor shall lead to new terms contributing to $I(x,t)$ (see App.\ref{ssec:front-shape}) which are proportional to $(\theta\Delta)^2$. Thanks to the fact that in our case $\theta\Delta\ll 1$ we can neglect them. However these extra terms has to be kept for larger $\Delta$, to ensure SU(2) invariance\footnote{the SU(2) invariance demands that the correlations of $\theta(x)$ and $\phi(x)$ fields have to be the same} at the Heisenberg point. [note that the singularity of $\theta_1(\Delta)$ when $\Delta\rightarrow 1$ can be interpreted as an effective increase of the $\theta$ coefficient].

The shift of the argument of all functions by $-(\theta\Delta)^2$ has
more non-trivial implications (we note that a full analytical solution with the given boundary conditions does exist and also have an argument shifted in the same way). Any observable, any correlation function $I(x,t)$, is a functional of a momentum integral of $F(q,t)$ as outlined in App.\ref{ssec:front-shape}. A standard procedure, implemented e.g. in Ref.\cite{Bernier14}, is to divide the whole range of integration into distinct ranges where the Bessel functions are either monotonous (for argument $<2/3$) or oscillating (for argument $>2/3$). Shifting argument of Bessel function effectively (up to terms of order $(\theta\Delta)^3$) shifts the ranges of integrals. This is particularly important for the intermediate range of momenta which describes intermediate distances, where the moving front is located. Instead of ranges defined as $[(2/3)\tilde{t}^{-3/2},2/3]$ now we have $[(2/3+\theta\Delta)\tilde{t}^{-3/2},2/3+\theta\Delta]$. The peculiar stretched exponential shape of the front remains but, an amplitude of the
intermediate regime is alerted. In general one can
deduce that it is the front that will be the most susceptible
to Galilean invariance breaking.

For larger values of $\Delta$, for $\Delta>0.5$, the linear approximation fails and one has to  resort to another approximation for the logarithmic derivative of $u(t)K(t)$.

\subsection{Large $\Delta$}

Probably the most interesting aspect is what happens with the mode occupation of TLL
when, while quenching, we approach the Heisenberg point where a
phase transition to the gapped phase takes place. It is reflected in Eq.\ref{eq:DGLa-XXZ} by the fact that the
$\theta_1$ term diverges. One can try to write down an approximate ODE
which would capture this effect and at the same time give a linear
increase of velocity. One way is to express Eq.\ref{eq:DGLa-XXZ} as a direct
generalization of the Euler equation:
\begin{equation}
(1-gt)^2\frac{\dd^2}{\dd t^2}a_n(t)+(1-gt)k_n\frac{\dd}{\dd t}
a_n(t)+(1-g^2t^2)^2 k_n^2 a_n(t)=0, \label{eq:DGLa-XXZ-Euler}
\end{equation}
Another is to take the approximation $\theta_{1}(t)=\Delta(t)/(1-\Delta(t))$.% can be quite well approximated by much simpler
%functions:
%\begin{equation}\label{eq:coeff-XXZ1-s}
%  \theta_{1}(k_n,t)= 1/(1-\Delta(t))
%\end{equation}
%\begin{equation}\label{eq:coeff-XXZ2-s}
%  \theta_{2}(k_n,t)= (\Delta(t)+1)^2
%\end{equation}
Both cases are analytically solvable and the solution is given in terms of the $HeunB$ functions. These novel functions are
very hard to operate with on the computational side\footnote{the class of Heun functions is not supported by \emph{Mathematica} only by \emph{Maple}} so we refrain from elaborating on the exact expressions that are rather lengthy. Nevertheless, it is interesting to note
that this class of functions is actually relevant as a solution of a well known problem in 1D physics.

To get an analytic insight into this critical regime one can try even stronger approximation, with $\theta_2=cste$, then the ODE is:
\begin{equation}\label{eq:XXZ-Heisen-ODE-large}
(\pi/2  J q)^2 F(t)-\frac{\theta }{1-\frac{\Delta  t}{\tau }}F'(t)+F''(t)=0
\end{equation}
The main advantage is that its solution can be again expressed in the form of a Bessel function which allows for a direct comparison with Ref.\cite{Bernier14}. The solution reads:
\begin{widetext}
\begin{equation}\label{eq:XXZ-Heisen-sol-large}
  F(t)= (2 \Delta(t-\tau/\Delta))^{\frac{1}{2}-\frac{\theta  \tau }{2 \Delta }} \left[ c_1 J_{\frac{1}{2}-\frac{\theta  \tau }{2 \Delta }}\left(J \pi/2 q( t-\frac{\tau }{\Delta})\right)+c_2 Y_{\frac{1}{2}-\frac{\theta  \tau }{2 \Delta }}\left(J \pi/2 q( t-\frac{\tau }{\Delta})\right)\right]
\end{equation}
\end{widetext}

We see that now the quench rate enters to the index of the Bessel
function. This is a highly non-perturbative effect. Based on the
derivation done in App.\ref{ssec:front-shape} we deduce that this
is able to affect the front shape, it is a stretched exponential
with an exponent that depends on the quench rate $\Delta/\tau$. We
emphasize that in our reasoning we assumed that the $\theta_1$
term proportional to $\partial_t \Delta(t)$ dominates the ODE, so
taking adiabatic limit $\tau\rightarrow\infty$ in
Eq.\ref{eq:XXZ-Heisen-sol-large} is not allowed. In the limit of
sudden quench, when $\tau\rightarrow 0$, the index of the Bessel
function goes to $1/2$. Implication for the front shape is that it
becomes more and more sharp, as the exponent of the stretched
exponential goes to zero. We also note that $J_{1/2}(z)\approx
\sin(z)/\sqrt{z}$, so in this limit our solution (expressed in
terms of trigonometric functions) is equal to an exact analytic
solution derived for the sudden quench in
Ref.\cite{Cazalilla06}.

\section{Beyond the linear ramp}

\subsection{Varying $\Delta(t)$} Based on Eq.\ref{eq:theta-depend} one see that the $\theta_1$ term of the ODE can be varied substantially if we change the quenching protocol. We then wish to construct the quench protocol such that the ODE will never run into any singularity at Heisenberg point. One can even propose cases where $\Delta(t)$ is finite, but $\partial_t \Delta(t)$ is zero. In one case of such protocol, $\Delta(t)=\Delta(-\frac{1}{2}(t/\tau)^2)-(t/\tau)$, the analytic solution exist and is discussed in the App.\ref{ssec:quadratic}.

\subsection{Constant $\theta_1$} In order to ensure that at every time of the quench the Galilean invariance breaking $\theta_1$ term in Eq.\ref{eq:DGLa-XXZ} is time independent (which greatly simplifies the ODE) we need to solve a following equation:
\begin{equation}\label{eq:diff-for-delta}
  \Delta '(t)=\frac{c \left(\Delta (t)-\Delta (t)^3\right)}{\Delta (t)^2+1}
\end{equation}
where $c$ is the constant value of $\theta_1$ term which we arbitrarily set. The Eq.\ref{eq:diff-for-delta} has a solution that reads:
\begin{equation}\label{eq:delta(t)}
  \Delta (t)=\Delta\frac{1}{2} e^{-c t} \left(\sqrt{4 e^{2 c t}+a^{2}}-a\right)
\end{equation}
where we have an arbitrary choice of $c_0$ and $a$ to build quenches of various profiles. It is possible to fix their ratio in a way such that $\Delta(t\rightarrow 0)\rightarrow 0$, but we keep it unconstrained, just demand $\Delta(t\rightarrow 0)\ll J$. %but they need to be fixed by initial and final $\Delta(t)$ and by demanding that $\Delta(t=0,\tau)$ is continuous. This has to be taken with initial conditions of ODE. The system of equations have too many constraints, which implies that the special protocol Eq.\ref{eq:delta(t)} is physically relevant only for one choice of initial and final $\Delta(t)$. This is contrary to all previous cases when a solution always existed for the boundary conditions of ODE.

%At this special point of initial/final $\Delta$
We have an ODE equation with constant $\theta_1$ and exponentially increasing velocity. The protocol described by general Eq.\ref{eq:delta(t)} is not analytically solvable, however it is enough to demand $c\gg a$, complete the square under the square-root and perform an appropriate Taylor expansion to arrive at a simpler exponential increase of velocity:
\begin{equation}\label{eq:expon-ODE}
(J q)^2 (1+ \Delta(1/2+a e^{-c t/\tau} )) F(t)- c \cdot F'(t)+F''(t)=0
\end{equation}
This equation is solvable:
\begin{widetext}
\begin{equation}\label{eq:exponential-XXZ}
F(t)=\left(\tau c J q \sqrt{a\Delta e^{-\frac{c t}{\tau }}}\right)^{-\frac{\tau }{2}} \left(c_2 \Gamma \left(\nu_3+1\right) J_{\nu_3}\left(\frac{2 \sqrt{a} \sqrt{e^{-\frac{c t}{\tau }}} J q \sqrt{\Delta } \tau }{c}\right)+c_1 \Gamma \left(1-\nu_3\right) J_{-\nu_3}\left(\frac{2 \sqrt{a} \sqrt{e^{-\frac{c t}{\tau }}} J q \sqrt{\Delta } \tau }{c}\right)\right)
\end{equation}
\end{widetext}

where a Bessel function of the first kind with an index
$\nu_3=\frac{\sqrt{c^2-4 J^2 q^2 (\Delta +1)} \tau }{c}$. The
solution Eq.\ref{eq:exponential-XXZ}, with an interaction
dependent index, also confirms the approximation we made to obtain
the solution in the Eq.\ref{eq:XXZ-Heisen-sol-large} where a
similar expression was found. It should be noted that, contrary to
the previous case, now the index of Bessel function does depend on
$q$, which makes an integration over $q$, necessary to obtain
correlation functions (see App.\ref{ssec:front-shape}), a much
more difficult task.

\section{Conclusions}

In conclusion, we have explored how the Galilean invariance breaking enters the dynamics of XXZ spin chain during the quench of its anisotropy $\Delta(t)$. We are able to obtain exact analytical solutions in several limiting cases. For the case of small $\Delta$ we can make a direct comparison of \emph{full analytic} solutions with and without the term breaking the Galilean invariance. We noticed that the key differences are: an extra exponential factor in correlation functions and a modified amplitude of the front. Upon increasing $\Delta$, hence approaching the Heisenberg point, another relatively simple analytical solution can be derived. Here we see that the quenching rate changes the shape of the front: it modifies the power of the stretched exponential that describes the front. Furthermore, we notice that by changing the quenching protocol, the time dependence $\Delta(t)$, we are able to suppress the term breaking the Galilean invariance, also in the vicinity of the Heisenberg point. This changes the bosonic modes occupation at intermediate momenta, which poses an intriguing question whether by changing the quenching protocol one may be able to influence the Renormalization Group flow and hence the physics of the non-equilibrium phase transition.

%\acknowledgments

\section*{Acknowledgments}

I would like to thank Dirk Schuricht for useful discussions in a very early stage of this work.

\appendix

\section{Approximate solution for intermediate $\Delta$}\label{ssec:quadratic}

Instead of linear ramp we now take the protocol for which $d \Delta(t)/ d t =0 $ at  $t=\tau$. An obvious choice is $\Delta(t)= \Delta(- t^2/2 +t)$, because then the coefficient of the Galilean invariance breaking $\theta_1$ term in Eq.\ref{eq:DGLa-XXZ} is approximately constant for $\Delta$ that is sufficiently large (but not at Heisenberg point). This gives the following ODE:

which is fortunately solvable. The solution reads:
%\begin{widetext}
\begin{equation}\label{eq:caseB-sol}
  F(t)=e^{\tilde{\nu}_e}{ c_1\cdot H_{-\tilde{\nu}}\left(\Upsilon_3\right)+c_2\cdot _1F_1\left(\tilde{\nu}/2;\frac{1}{2};\left(\Upsilon_3\right)^2\right)}
\end{equation}
%\end{widetext}
where the index $\tilde{\nu}=\frac{\theta ^2 \tau -6 J^2 q^2 \tau +2 \sqrt{2} \Delta  J q}{4 \sqrt{2} \Delta  J q}$, $\tilde{\nu}_e=\frac{t}{2 \tau }(-\frac{\Delta  J q t}{\sqrt{2}}+\sqrt{2} J q \tau +\theta\tau)$ and the variable $\Upsilon_3 = \frac{\sqrt{\Delta } \sqrt{J} \sqrt{q} t}{\sqrt[4]{2} \sqrt{\tau }}-\frac{\sqrt{J} \sqrt{q} \sqrt{\tau }}{\sqrt[4]{2} \sqrt{\Delta }} $. This functional form cannot be converted to Bessel functions. From the fact that argument of an exponential depends on q we expect that effective cut-off of the theory will be time dependent, moreover the correction terms to $\theta(x)$ correlation functions that comes from a derivative $\partial_t F(q,t)$ can now be significant especially at large $q$ (there is no $\theta\Delta$ factor that would diminish them).

\section{Shape of the front}\label{ssec:front-shape}

%The limit of small (and constant) time is defined by a condition $Cos(v q t)\approx 1$. Obviously this condition is well suited to look more carefully at limit $q\approx 0$, which according to the stationary point approximation turns out to be crucial. From Eq.\ref{eq:after-sol-gen} it is clear that here solutions for
%$v(q,\tau)$ obtained from ODE during the quench can be directly applied. For concreteness we begin with a study of $I_{\theta\theta}(x,t)$ for the linear ramp protocol.

To compute a correlation function e.g. $\langle\theta(x,t)\theta(0,t)\rangle$ we need to evaluate the following integral:
\begin{equation}\label{eq:q-integral}
    I(x,t)=\langle\theta(x,t)\theta(0,t)\rangle= \int \frac{dq}{q^3}\exp(-\tilde{\alpha}|q|) \sin(q x/2))^2 |F(t)|^2
\end{equation}
where $\tilde{\alpha}$ is a UV cutt-off of the theory. In case of
Eq.\ref{eq:XXZ-linear-Bessel} or Eq.\ref{eq:XXZ-Heisen-sol-large},
the formula Eq.\ref{eq:q-integral}  boils down to an integral over
a combination of Bessel functions. Following standard procedure we
divide Bessel function into a monotonous part power law
$J_\nu(z)\sim 1/z^{\nu}$ (for small argument $z<2/3$) and an
oscillating part (for large argument $z>2/3$). The most important
range of integration, that determines the shape of propagating
front is the intermediate range of momenta defined in
Ref.\cite{Bernier14}: time dependent Bessel functions are
monotonous while the time-independent amplitudes (set by the
boundary conditions) Bessel functions oscillates.
%We get incomplete gamma function $\Gamma(1/3,x)$ with the
%following asymptotic behaviour:
%\begin{eqnarray}
% \nonumber to remove numbering (before each equation)
%  \Gamma(1/3,x) &\rightarrow& log(x)~~(for~k_c????) \\
%  \Gamma(1/3,x) &\rightarrow& x^{1/3}\Gamma(1/3)~~(for~k_c????)
%\end{eqnarray}\label{eq:front-shape}
%which for small ramp speed falls of like a $log(x)$ and thus gives
%rise to power law. Where the second limit stems from the fact that
%for large quench rates the upper limit of the integral can be
%neglected and we are left with a complete gamma function times the
%$x^{1/3}$ (for linear ramp). , while the second expression leads
%to a transitional, stretched exponential dependence.

\subsection{Bessel function $J_{\pm 1/3}(q,t)$}

In this case the most divergent term of Eq.\ref{eq:q-integral} in $q\rightarrow 0$ limit takes the following form in this regime\cite{Bernier14}:
\begin{equation}\label{eq:q0domin}
  I_{fro}(x,t)\approx\int_{b_1}^{b_2} dh_q \frac{(\sin(h_q \tilde{t}_q)\sin(q x/2))^2}{h_q^{4/3}}
\end{equation}
where, from Eq.\ref{eq:XXZ-linear-Bessel}
$\tilde{t}_q=\Upsilon_2/q^2$  and $h_q=
q/(12{J}^{2}{\tau}^{2}\Delta)$. The first sine stems from the
(oscillating) approximation of Bessel function and the second is
from the Fourier transform. The limits of integral are: $b_1= 2/3
\tilde{t}_q^{-3/2}$ and $b_2=2/3$. The very existence of this
intermediate regime, $b_1<b_2$ is related to $\tilde{t}_q>1$. This
is possible only for substantial amplitude of a quench, which
clearly indicates that we study effects inaccessible by any
perturbative approach. The integration in Eq.\ref{eq:q0domin} can
be performed in a closed analytical form, it leads to:
\begin{equation}\label{eq:result-t-tau}
    I_{fro}(x,t)= z_{0}^{1/3} \Gamma(2/3,z_0)|_{b_1}^{b_2}
\end{equation}

where $\Gamma(2/3,z)$ is the incomplete Gamma function
and $z_0=2\imath q(vt-x)$. We first estimate an upper limit term, $\Gamma(2/3,2\imath b_2 (vt-x))$. We use the identities
$\Gamma(s, z)=(1-s)\Gamma(s-1,z)+\exp(z)z^{s-1}$ and
$\Gamma(s,z)=z^{s} E_{1-s}(z)$ where $E_{1-s}(z)$ is the
generalized exponential function to obtain
$$
\Gamma(2/3,z)=z^{-1/3}(-\frac{1}{3}E_{4/3}(z)+\exp(\imath z))
$$
and the $z^{-1/3}$ above cancels out with $z^{1/3}$ in
Eq.\ref{eq:result-t-tau} and one is left with the $E_{4/3}(z)$
which in turn for large $(vt-x)$ ($q=b_2$ is kept constant) can be approximated well by the $Log[z]$ times an oscillatory function
(extracting the oscillating part leads to the Triconi confluent
hypergeometric function $U(4/3,4/3,z)$ ). We have arrived at the desired $I_{fro}(x,t)\sim Log(x)$ behaviour. We see that for large $(vt-x)$ we make a smooth cross-over to a power law behaviour
expected deep inside the light cone (it matches well with the
\emph{adiabatic range}).

For intermediate $z$, that is small to moderate $(vt-x)$, there exist a range where $\Gamma(s,z)\approx 0$ thus a term originating from upper integration limit is negligible
and it is the term originating from lower integration limit $b_1$
that dominates. When $b_1$ goes to zero the
incomplete Gamma function remains finite (actually it even
increases) and it equals to the complete Gamma function
$\Gamma(2/3)$. The $z^{1/3}$ does not drop out and this is the source
of stretched exponential behavior discovered in Ref.\cite{Bernier14}: $I_{fro}(x,t)\sim \Gamma(2/3)\exp[x^{1/3}]$.

\subsection{Bessel function with arbitrary index}

Once we have shown how to re-obtain results of Ref.\cite{Bernier14} we can generalize them to Bessel function with an arbitrary, but momentum independent index. What is changing is the exponent in the denominator of Eq.\ref{eq:q0domin} which follows the law $1+a=2(1-\nu_3)$. This implies the following functional form of the result (analogue of Eq.\ref{eq:result-t-tau} for general $\nu$):
\begin{equation}\label{eq:q0domin-powers}
 I_{fro}(x,t)\approx\int_{b_1}^{b_2} dq \frac{(\sin(h_q \tilde{t}))\sin[qx]^2}{q^{1+a}}= z_{0}^{a} \Gamma(1-a,z_0)|_{b_1}^{b_2}
\end{equation}

where now $h_q= J \pi/2 q$ and $\tilde{t}= t-\frac{\tau }{\Delta}$. One can perform exactly the same manipulations like before to arrive at the stretched exponential shape of the front, but now the exponent is modified $I_{fro}(x,t)\sim \exp[x^{a}]$. For $\epsilon\ll 1$ the $\Gamma(-a,\imath \epsilon)$ is an increasing function of $a$. As a result the amplitude of the stretched exponential region becomes larger as $\nu$ decreases. For a special case $a=1/2$ the result can be also rewritten in terms of Fresnel integral $C(z)$.

\bibliography{biblioXXZ}
%\bibliography{paper}

\end{document}